\documentclass[aoas,nameyear,dvips]{arximspdf}
\usepackage{dcolumn}
\usepackage{graphicx}

\doi{10.1214/10-AOAS371}
\volume{5}
\issue{1}
\pubyear{2011}
\firstpage{176}
\lastpage{200}

\makeatletter

\newcolumntype{d}[1]{D{.}{.}{#1}}

\newtheorem{tm}{Theorem}
\newtheorem{la}{Lemma}
\newtheorem{cy}{Corollary}

\newproclaim{exper}{Experiment}
\newproclaim{remark}{Remark}

\makeatother

\begin{document}
\begin{frontmatter}

\title{Estimating the number of neurons in multi-neuronal spike
trains\thanksref{TITL1}}
\thankstext{TITL1}{Supported in part by National University of
Singapore Grant R-155-000-094-112.}
\runtitle{Multi-neuronal spike train}

\begin{aug}
\author[A]{\fnms{Mengxin} \snm{Li}\ead[label=e1]{doadvan@gmail.com}}
and
\author[B]{\fnms{Wei-Liem} \snm{Loh}\corref{}\ead[label=e2]{stalohwl@nus.edu.sg}}
\runauthor{M. Li and W.-L. Loh}
\affiliation{National University of Singapore and National University
of Singapore}

\dedicated{Dedicated to Charles M. Stein on his ninetieth birthday}

\address[A]{Genome Institute of Singapore\\
60 Biopolis Street\\
\#02-01, Genome\\
Singapore 138672\\
Republic of Singapore\\
\printead{e1}} 

\address[B]{Department of Statistics\\
and Applied Probability\\
National University of Singapore\\
Singapore 117543\\
Republic of Singapore\\
\printead{e2}}
\end{aug}

\received{\smonth{10} \syear{2009}}
\revised{\smonth{6} \syear{2010}}

%
\begin{abstract}
A common way of studying the relationship between neural activity and
behavior is through the analysis of neuronal spike trains that are
recorded using one or more electrodes implanted in the brain. Each
spike train typically contains spikes generated by multiple neurons.
A natural question that arises is ``what is the number of neurons $\nu$
generating the spike train?''
This article proposes a method-of-moments technique for estimating $\nu$.
This technique estimates the noise nonparametrically using data from
the silent region of the spike train and
it applies to isolated spikes with a possibly small, but nonnegligible,
presence of overlapping spikes.
Conditions are established in which the resulting estimator for $\nu$
is shown to be strongly consistent.
To gauge its finite sample performance,
the technique is applied to simulated spike trains as well as to actual
neuronal spike train data.
\end{abstract}

%

\end{frontmatter}

\section{Introduction}\label{sec1}

In the field of neuroscience, it is generally  acknowledged that
neurons are the basic units of information processing in the brain.
They play this role by generating highly peaked electric action
potentials or, more simply, spikes [cf.\ Brillinger (\citeyear{B1988}),
Dayan and Abbott (\citeyear{DA2001})]. A sequence of such spikes over
time is called a spike train. A typical method of recording spike
trains is by inserting electrodes into the brain. In the analysis of a
spike train, Brown, Kass and Mitra (\citeyear{BKM2004}) note three
goals: (i) identify each spike as ``signal'' (versus pure noise), (ii)
determine the number of neurons being recorded, and (iii) assign each
spike to the neuron(s) that produced it. (i), (ii) and (iii) are
collectively termed \textit{spike sorting}
 in the neuroscience literature, and, as remarked by Brown, Kass and
Mitra (\citeyear{BKM2004}), are mandatory for all multi-neuronal spike
train data analyses. The accuracy of spike sorting critically affects
the accuracy of all subsequent analyses. For spike sorting to be
well defined, it is generally assumed that each neuron generates a
spike with a characteristic voltage waveshape (apart from noise) and
that distinct neurons have distinct spike waveshapes. For example,
Figure \ref{fig:2} in Section~\ref{sec4} presents the different spike waveshapes
of 5  neurons that were estimated from real data.

This article assumes that (i) has been achieved, that is, each spike in
the spike train has been identified. The objective here is to estimate
the number of neurons that produced the spike train using the data from
(i). This problem has the following feature that makes it a nonstandard
clustering problem. Most often, a spike is produced by one (and only
one) neuron, as no other neuron spikes around that time. Such a spike
is called an \textit{isolated spike}. On the other hand, there may be
instances where two or more neurons spike in synchrony, that is,\ at
almost the same time [cf.\ Lewicki (\citeyear{L1998}), page R68]. Then the
resulting spike is an additive superposition of the spikes generated by
this group of neurons. This spike is called an \textit{overlapping
spike}. Consequently, if considered as a clustering problem, the number
of clusters will not in general be equal to the number of neurons
producing the spike train.

For definiteness we shall assume that there are $\nu$ neurons, labeled
$1$ to $\nu$, generating the spike train and that $n$ spikes, say,
$S_1,\ldots, S_n$, are detected and recorded in the spike train. Here
the $S_i$'s are aligned according to their peaks and each $S_i\in
{\mathbb R}^d$ for some $d\in {\mathbb Z}^+$. The $d$ components of
$S_i$ are the measurements of the voltage values of the spike on a
regular grid of time-points around the peak (or maximum) of the spike.
Writing $S_i = (S_{i,1},\ldots, S_{i,d})' \in {\mathbb R}^d$, we assume
that
\begin{equation}
S_{i,j} = \Theta_{i,j} + \eta_{i,j}\qquad   \forall i=1,\ldots, n, j=1,\ldots, d,
\label{eq:1.7}
\end{equation}
where $\eta_{i,j}, i=1,\ldots, n, j=1,\ldots, d$, are i.i.d.\ noise
random variables with mean 0, variance $\sigma^2$ and $\Theta_i=
(\Theta_{i,1},\ldots, \Theta_{i,d})' \in {\mathbb R}^d$, $i=1,\ldots,
n$, are i.i.d.\ random vectors. The $\Theta_i$'s and $\eta_{i,j}$'s are
assumed to be all independent. $\Theta_i$ denotes the denoised spike
shape of $S_i$ and is random because the denoised spike shape is a
function of the particular neuron(s) generating it and time lag in the
spiking times if there are $\geq $2 neurons. Let $\alpha \in {\mathbb
R}^d$ be a constant vector such that $\alpha' \alpha =1$. In Sections \ref{sec4}
and \ref{sec5}, $\alpha$ is taken to be the first principal component of
$S_1,\ldots, S_n$ and $0.01 n$ vector $0$'s. However, other choices of
$\alpha$ are possible too. For each spike $S_i$, define
\begin{equation}
X_i = \alpha' S_i\qquad  \forall i=1,\ldots, n,
\label{eq:1.9}
\end{equation}
 which is the projection of $S_i$ onto $\alpha$.
It follows from (\ref{eq:1.7}) that  $X_1,\ldots, X_n$ are i.i.d.\ random variables.
As observed by Ventura (\citeyear{V2009}), either implicitly or explicitly, almost all spike sorting methods assume that $X_i$
has a mixture distribution
with probability density function of the form
\begin{equation}
\sum_{q=1}^\nu \sum_{1\leq j_1<\cdots < j_q \leq \nu} \pi_{j_1,\ldots, j_q} h_{j_1,\ldots, j_q} (x)\qquad  \forall x\in \mathbb R,
\label{eq:1.1}
\end{equation}
where $\pi_{j_1,\ldots, j_q}$ is the probability that $S_i$ is
generated by (and only by) neurons $j_1,\ldots, j_q$. The
$h_{j_1,\ldots, j_q}$'s are usually assumed to be Gaussian densities
[cf.\ Lewicki (\citeyear{L1994}, \citeyear{L1998})], even though $t$ densities have also been
proposed [cf. Shoham, Fellows and Normann (\citeyear{SFN2003})].

In this article we shall not assume that $h_j(\cdot)$, $j=1,\ldots, \nu$, are Gaussian or  $t$ densities but only that
$h_j(\cdot) = f_{\mu_j,\sigma^2} (\cdot)$ belongs to a location family of probability densities
[cf. Lehmann (\citeyear{L1983}), page 154].
It is noted that this location family contains both Gaussian and $t$ densities.
Here $\mu_j$ and $\sigma^2$ denote the mean and variance induced by the density $f_{\mu_j, \sigma^2}$.

Due to the complex nature of spike overlap, we do not think it is accurate to model each $h_{j_1,\ldots, j_q}$, $2\leq q\leq \nu$,
as a Gaussian or a $t$ density (or, indeed, any other parametric density having only a small number of unknown parameters).
For example, $h_{1,2}$
depends on noise as well as on the time lag between the spiking times of neurons 1 and 2. As the time lag is also random and
that the phenomenon of neurons spiking in close proximity to each other is still not well understood, we feel
it is more appropriate to model $h_{1,2}$ using a nonparametric density (rather than a parametric one).
In particular, in this article, we shall assume that (\ref{eq:1.1}) is of the form
\begin{eqnarray}\label{eq:1.2}
f( x) &=& \sum_{j=1}^\nu \pi_j f_{\mu_j, \sigma^2}
(x)\nonumber\\[-8pt]\\[-8pt]
&&{}+ \sum_{q=2}^\nu \sum_{1\leq j_1<\cdots < j_q\leq \nu} \pi_{j_1,\ldots, j_q} \int_{\mathbb R} f_{\mu, \sigma^2} (x) \,d G_{j_1,\ldots, j_q} (\mu),\nonumber
\end{eqnarray}
where $\pi_1 \geq \pi_2 \geq \cdots \geq \pi_\nu > 0$ and $G_{j_1,\ldots, j_q}$'s are unknown absolutely continuous probability distributions.
Consequently, (\ref{eq:1.2}) leads to a nonparametric location mixture.

The objective is to estimate $\nu$ in (\ref{eq:1.2}) using the sample $X_1,\ldots, X_n$ and an independent auxiliary sample
of i.i.d.\ observations $Y_1,\ldots, Y_m$ obtained from the silent region of the spike train.
The silent region is defined to be the sections of the spike train where there are no spikes.
Hence, the $Y_l$'s are defined as
\begin{equation}
Y_l = (\eta_{l,1}^*, \ldots, \eta_{l,d}^*) \alpha\qquad    \forall l=1,\ldots, m,
\label{eq:1.8}
\end{equation}
where $\eta^*_{l,1},\ldots, \eta^*_{l,d}$ are noise voltage measurements on a regular grid of $d$ consecutive time-points of the silent region of the spike train.
We assume that $\eta_{l,j}^*, \eta_{i,k}$, $1\leq l \leq m,  1\leq i\leq n, 1\leq j, k\leq d$, are i.i.d.\ random variables with mean 0 and variance $\sigma^2$.
In this article a method-of-moments estimator $\hat{\nu}$ for $\nu$ is proposed.
This estimator has a number of attractive properties. First, this article establishes a reasonably transparent theory justifying/supporting $\hat{\nu}$.
In particular, $\hat{\nu}$ is a strongly consistent estimator for $\nu$ under mild conditions.
Second, the estimator can be computed without first (or concurrently) computing the other unknown quantities in (\ref{eq:1.2}).
Consequently, it is computationally very fast relative to, say, EM or Markov chain Monte Carlo (MCMC) algorithms.

We would like to add that the above problem of estimating $\nu$ can be
regarded as robust and (yet) consistent estimation of the number of
components of a finite mixture where the latter is subjected to a small
but nonnegligible contamination by a nuisance distribution. While
quite a number of papers have been written on estimating the number of
components of a finite mixture [cf. Dacunha-Castelle and Gassiat
(\citeyear{DCG1997}) and references cited therein], we are not aware of any work in
the statistics literature that deals with this problem when the mixture
is contaminated by another distribution.

On the neuroscience literature side, numerous algorithms for spike
sorting have been proposed. A review of spike sorting algorithms and a
discussion of their strengths and limitations can be found in Lewicki
(\citeyear{L1998}). In particular, a considerable number of spike
sorting algorithms assume that the proportion of overlapping spikes is
negligible relative to the proportion of isolated spikes and, hence,
the possible presence of overlapping spikes is ignored. With this
assumption, many spike sorting algorithms further assume that the
number of neurons $\nu$ is known and the problem reduces to a standard
classification problem. If $\nu$ is unknown, other spike sorting
algorithms use various EM or MCMC methods for determining $\nu$ as well
as for assigning spikes to the neurons [cf.\ Pouzat, Mazor and Laurent
(\citeyear{PML2002}), Nguyen, Frank and Brown (\citeyear{NFB2003}), Wood and Black (\citeyear{WB2008})].
However, Brown, Kass and Mitra [(\citeyear{BKM2004}), page 456] noted
that MCMC techniques have yet to be widely tested in spike sorting.

Spike sorting algorithms that take overlapping spikes into account
usually involve significant user input [cf.\ Mokri and Yen (\citeyear{MY2008})].
Section 5 of Lewicki (\citeyear{L1998}) discusses the use of templates, independent
component analysis  and neural networks to handle overlapping spikes
and the limitations of these methods. If $\nu$ is known, there are at
least two model-based approaches for handling overlapping spikes. The
first approach considers a $(\nu+1)$-component mixture distribution
with the first $\nu$ components modeling the spike waveforms of the
$\nu$ neurons. The $(\nu +1)$th component is a uniform density over a
suitably large region of the sample space [cf.\ Sahani (\citeyear{S1999}), page 95]
that serves as an approximate model for the overlapping spikes. The
second approach is a trimming method where a number of the largest
observations (outliers) are omitted [cf.\ Gallegos and Ritter (\citeyear{GR2005}),
Garc\'{i}a-Escudero et al. (\citeyear{GEetal2008})]. The rationale is that
most of the outliers correspond to overlapping spikes and, hence, the
remaining observations should be comprised  essentially of isolated
spikes.

\begin{table}[t]
\caption{Factorial experimental design}\label{table:3.1.0}
\begin{tabular}{@{}lccc@{}}
\hline
\textbf{Experiment} & \textbf{Noise distribution} & \textbf{Spike detection algorithm} & \textbf{Proportion of}\\
&&&\textbf{overlapping spikes}\\
\hline
 1 & Gaussian& Oracle& 10\%\\
 2 & Gaussian& SpikeOMatic& 10\%\\
3 & Gaussian& Oracle& 0\\
 4 & Gaussian& SpikeOMatic& 0\\
 5 & Student $t_5$ & Oracle& 10\%\\
 6 & Student $t_5$ & SpikeOMatic& 10\%\\
 7 & Student $t_5$ & Oracle& 0\\
 8 & Student $t_5$ & SpikeOMatic& 0\\
\hline
\end{tabular}
\end{table}

The rest of the article is organized as follows.
Section \ref{sec2} introduces a number of trigonometric moment matrices. Theorem \ref{tm:2.1} derives some explicit error bounds for their eigenvalues.
Motivated by these error bounds, Section \ref{sec3} proposes a method-of-moments estimator $\hat{\nu}$ for $\nu$.
Theorem \ref{tm:3.2} shows that $\hat{\nu}$ is a strongly consistent estimator for $\nu$ under mild conditions.
A point of note is that Theorem \ref{tm:3.2} does not require the proportion of overlapping spikes to be asymptotically negligible
 as sample size tends to infinity.

Section \ref{sec4} presents a detailed simulated spike train study that
investigates the finite sample accuracy of $\hat{\nu}$. Each spike
train is generated by $\nu$ neurons where $\nu$ varies from 1 to 5. The
spike shapes of these neurons are estimated from real data. There are 8
experiments in the study which present a variety of different spike
train situations depending on the proportion of overlapping spikes, the
spike detection technique and  sample sizes $n, m$. Table
\ref{table:3.1.0} summarizes the 8 experiments as a factorial design.
This study finds that $\hat{\nu}$ has very good accuracy with regard to
these 8 experiments for moderately large sample sizes such as
$n=1000$ and $m=2000$.
 As a comparison, we have also applied the SpikeOMatic software [cf.\
Pouzat, Mazor and Laurent (\citeyear{PML2002}); Pouzat et al.
(\citeyear{Petal2004})] to obtain an alternative estimate $\hat{\nu}_1$
for $\nu$ in 2 of these experiments.

\begin{figure}

\includegraphics{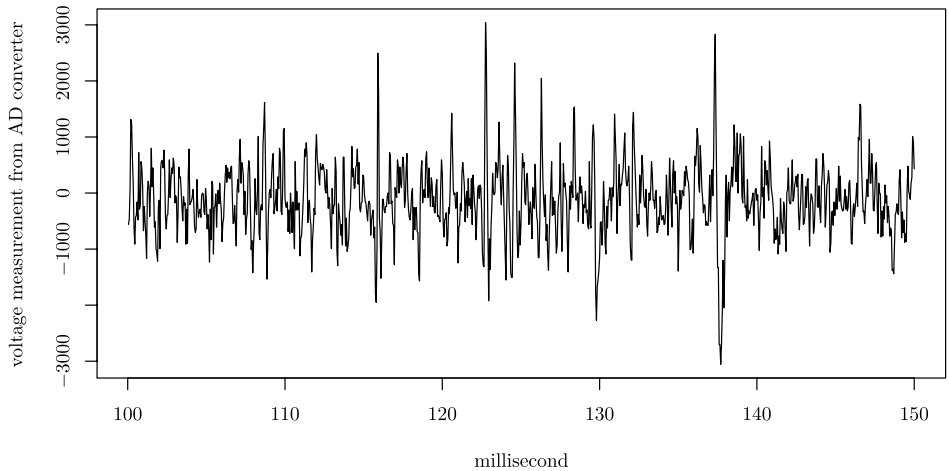}

  \caption{A portion of Lewicki's actual spike train recording.}
  \label{fig:0}
\end{figure}

Section \ref{sec5} considers two spike train data sets taken from
Lewicki (\citeyear{L1994}). The first is an actual 40-second spike
train recording and the second is a synthesized recording using 6 spike
shapes estimated from the first data set. Figure \ref{fig:0} presents a
portion of the actual spike train recording. Lewicki (\citeyear{L1994})
inferred that $\nu$ is 6 for the actual recording. Three estimators are
used to estimate $\nu$ from each spike train, namely, $\hat{\nu}_2$ by
Lewicki's spike sorting algorithm [cf.\ Lewicki (\citeyear{L1994})],
$\hat{\nu}_1$ by SpikeOMatic software [cf.\ Pouzat, Mazor and Laurent (\citeyear{PML2002})] and $\hat{\nu}$. For the synthesized recording, we obtain
$\hat{\nu}_1 = \hat{\nu}_2 = 5$ while $\hat{\nu} = 4$ or $5$ depending
on whether the threshold of $\hat{\nu}$ is set to 1.0 or 0.8,
respectively. On the other hand, for the actual spike train recording,
we obtain $\hat{\nu}_1 = 12$, $\hat{\nu}_2 = 9$ while $\hat{\nu} = 4$
or $5$ depending on whether the threshold of $\hat{\nu}$ is set to 1.0
or 0.8, respectively. Thus, relative to $\hat{\nu}_1$ and $\hat{\nu}_2$,
$\hat{\nu}$ is a more stable estimate with respect to these 2 data
sets.

The article ends with some concluding remarks in Section \ref{sec6}.
The symbols ${\mathbb R}, {\mathbb C}$ denote the set of real numbers,
complex numbers respectively, and all proofs  are deferred to the
\hyperref[app]{Appendix}.

\section{Some trigonometric moment matrices}\label{sec2}

This section constructs a number of trigonometric moment matrices and derives explicit error bounds for their eigenvalues.
Following the notation in (\ref{eq:1.2}), let $\theta$ denote a random variable with cumulative distribution function
\begin{eqnarray}\label{eq:2.2}
F_\theta (\mu) = \sum_{j=1}^\nu \pi_j {\mathcal{I}}\{\mu \geq
\mu_j\}+ \sum_{q=2}^\nu \sum_{1\leq j_1 <\cdots < j_q\leq \nu} \pi_{j_1,\ldots, j_q} G_{j_1,\ldots, j_q}(\mu)\nonumber\\[-8pt]\\[-8pt]
\eqntext{\forall \mu\in \mathbb R,}
\end{eqnarray}
where $\pi_1 \geq \pi_2 \geq \cdots \geq \pi_\nu >0$ and ${\mathcal{I}}\{\cdot\}$ denotes the indicator function. For simplicity, let
$\pi_{\mathrm{cont}}$ be the proportion of overlapping spikes and $\pi_{\mathrm{cont}} F_{\mathrm{cont}}$ be the continuous component of $F_\theta$. Then
\begin{eqnarray*}
\pi_{\mathrm{cont}} &=& \sum_{q=2}^\nu \sum_{1\leq j_1<\cdots <j_q\leq \nu} \pi_{j_1,\ldots, j_q} = 1 - \sum_{j=1}^\nu \pi_j, \\
\pi_{\mathrm{cont}} F_{\mathrm{cont}} (\mu) &=& \sum_{q=2}^\nu \sum_{1\leq j_1 <\cdots < j_q\leq \nu} \pi_{j_1,\ldots, j_q} G_{j_1,\ldots, j_q}(\mu)\qquad  \forall \mu\in \mathbb R.
\end{eqnarray*}
Let $p\geq \nu$ be an integer.
Motivated by the ideas of Lindsay (\citeyear{L1989a}), (\citeyear{L1989b}) on moment matrices and finite mixtures, we define a function
$T_p \dvtx {\mathbb R} \rightarrow {\mathbb C}^{(p+1)\times (p+1)}$ by
\begin{eqnarray*}
T_p (x) &=&  \pmatrix{
1 & e^{\mathbf{i}x} &  e^{\mathbf{i}2x} & \ldots & e^{\mathbf{i}p x}\cr
e^{-\mathbf{i}x} & 1 & e^{\mathbf{i}x} & \ldots &
e^{\mathbf{i}(p-1)x}\cr
e^{-\mathbf{i}2x} & e^{-\mathbf{i}x} & 1 & \ldots &
e^{\mathbf{i}(p-2)x}\cr
\vdots & \vdots & \vdots & \ddots & \vdots\cr
e^{-\mathbf{i}px} & e^{-\mathbf{i}(p-1)x} & e^{-\mathbf{i}(p-2)x} & \ldots & 1
}
\\
&=& \pmatrix{1 \cr e^{-\mathbf{i}x} \cr e^{-\mathbf{i}2x} \cr \vdots \cr e^{-\mathbf{i}p x}} (1,e^{\mathbf{i}x}, e^{\mathbf{i}2x},\ldots, e^{\mathbf{i}p x}),
\end{eqnarray*}
where $\mathbf{i} = \sqrt{-1}$. We further define matrices corresponding to the discrete component and the continuous component of $F_\theta$ by
\begin{eqnarray}\label{eq:2.1}
M_{p, \mathrm{disc}}
&=& \sum_{i=1}^\nu \pi_i T_p(\mu_i)
=
 \pmatrix{
1 & \ldots & 1\cr
e^{-\mathbf{i}\mu_1} & \ldots & e^{-\mathbf{i}\mu_\nu}
 \cr
e^{-\mathbf{i}2\mu_1}& \ldots & e^{-\mathbf{i}2\mu_\nu}
 \cr
\vdots & \ddots & \vdots
 \cr
e^{-\mathbf{i}p\mu_1} & \ldots & e^{-\mathbf{i}p\mu_\nu}
}
\pmatrix{
\pi_1 & 0 &  \ldots & 0
\cr
0 & \pi_2 & \ldots & 0
 \cr
\vdots & \vdots & \ddots & \vdots
 \cr
0 & 0 & \ldots & \pi_\nu
}
\nonumber\\
&&\hphantom{\sum_{i=1}^\nu \pi_i T_p(\mu_i)
=}  {}\times
\pmatrix{
1 & e^{\mathbf{i}\mu_1} & e^{\mathbf{i}2\mu_1} & \ldots & e^{\mathbf{i}p\mu_1}
 \cr
\vdots & \vdots & \vdots & \ddots & \vdots
 \cr
1 & e^{\mathbf{i}\mu_\nu} & e^{\mathbf{i}2\mu_\nu} & \ldots & e^{\mathbf{i}p\mu_\nu}
},
 \\
M_{p, \mathrm{cont}} &= & \int_{\mathbb R}   T_p(\mu) f_{\mathrm{cont}} (\mu)  \,d(\mu) = \int_0^{2\pi} T_p(\mu) \sum_{j=-\infty}^\infty f_{\mathrm{cont}} (\mu+2\pi j) \,d(\mu),
\nonumber
\end{eqnarray}
where $f_{\mathrm{cont}}$ is the probability density function of the distribution $F_{\mathrm{cont}}$.
Finally, we define
\[
M_p = E [ T_p (\theta) ] = M_{p, \mathrm{disc}} + \pi_{\mathrm{cont}} M_{p, \mathrm{cont}}.
\]
Let $\lambda_i (A)$ denote the $i$th largest eigenvalue of $A$ where $A$ is an arbitrary $(p+1)\times (p+1)$ Hermitian matrix.
Hence, $\lambda_1 (A) \geq \lambda_2 (A)\geq \cdots \geq \lambda_{p+1} (A)$.

\begin{tm} \label{tm:2.1}
With the above notation, suppose that $\mu_1,\ldots, \mu_\nu$ are all distinct, $0\leq \mu_1, \ldots, \mu_\nu$ $< 2\pi$.
Then for $i=1,\ldots, \nu$, we have
\begin{eqnarray}\label{eq:2.3}
 && (p+1) \pi_i + 2\pi\pi_{\mathrm{cont}} \Biggl\{ \min_{0\le\mu < 2\pi} \sum_{j=-\infty}^\infty f_{\mathrm{cont}} (\mu+2\pi j) \Biggr\}\nonumber\\
 &&\quad {} -\sqrt{2\sum_{1\le j<k\le\nu}\pi_j\pi_k  \biggl|\frac{1-e^{\mathbf{i}(p+1)(\mu_j-\mu_k)}}{1-e^{\mathbf{i}(\mu_j-\mu_k)}} \biggr|^2 }
\nonumber\\
& &\qquad \leq  \lambda_i( M_p)
\\
& &\qquad \leq (p+1) \pi_i + 2\pi\pi_{\mathrm{cont}} \Biggl\{ \max_{0\leq \mu < 2\pi} \sum_{j=-\infty}^\infty f_{\mathrm{cont}} (\mu+2\pi j)\Biggr\}\nonumber\\
&&\qquad \quad {} +\sqrt{2\sum_{1\le j<k\le\nu}\pi_j\pi_k  \biggl|\frac{1-e^{\mathbf{i}(p+1)(\mu_j-\mu_k)}}{1-e^{\mathbf{i}(\mu_j-\mu_k)}} \biggr|^2 }.
\nonumber
\end{eqnarray}
Also, for $i=\nu+1,\ldots, p+1$, we have
\begin{eqnarray}\label{eq:2.4}
&&2\pi\pi_{\mathrm{cont}} \min_{0\le\mu < 2\pi} \sum_{j=-\infty}^\infty f_{\mathrm{cont}} (\mu+2\pi j) \nonumber\\[-8pt]\\[-8pt]
&&\qquad \le \lambda_{i}( M_p)\le 2\pi\pi_{\mathrm{cont}} \max_{0\le\mu < 2\pi} \sum_{j=-\infty}^\infty f_{\mathrm{cont}} (\mu+2\pi j).\nonumber
\end{eqnarray}
\end{tm}

The following is an immediate corollary of Theorem \ref{tm:2.1}.
\begin{cy} \label{cy:2.1}
Suppose the conditions of Theorem \ref{tm:2.1} are satisfied.
Then for any constant $\gamma>0$, there exists a positive integer $p_\gamma$ such that
\[
\lambda_\nu (M_p) > \gamma \sqrt{p+1} > \lambda_{\nu+1} (M_p)\qquad   \forall p\geq p_\gamma.
\]
Also, $\lambda_{\nu+1}(M_p)$ is bounded uniformly in $p$. Finally, $\lambda_i (M_p) \sim (p+1)\pi_i$, $\forall 1\leq i\leq \nu$, as $p\rightarrow \infty$.
\end{cy}

Corollary \ref{cy:2.1} gives, at least in principle, a way for estimating $\nu$ by estimating the eigenvalues of $M_p$.

\section{A method-of-moments estimator for $\nu$}\label{sec3}

Let $\eta_{i,j}$, $i=1,\ldots, n, j=1,\ldots, d$,
and $\Theta_i$, $i=1,\ldots, n$, be as in (\ref{eq:1.7}) and
$X_i = \alpha' S_i$, $i=1,\ldots, n$, be as in (\ref{eq:1.9}).
Then  $X_1,\ldots, X_n$ is an i.i.d.\ sequence of observations from the mixture distribution given by (\ref{eq:1.2}).
We observe that
\[
X_i = \theta_i + \tilde{Y}_i\qquad    \forall i=1,\ldots, n,
\]
where $\theta_i = \alpha' \Theta_i$ and $\tilde{Y}_i = (\eta_{i,1}, \ldots, \eta_{i, d})' \alpha$
 are independent
random variables having cumulative distribution function $F_\theta$, given by (\ref{eq:2.2}), and probability density function $f_{0,\sigma^2}$, respectively.
$\theta_i$ can be regarded as the signal and $\tilde{Y}_i$\vspace*{-3pt} the zero-mean noise.
Here we assume that
$E (e^{- \mathbf{i} k  \tilde{Y}_1}) \neq 0$ for all $k\in \{ 1, \ldots, p\}$.
This is a very weak assumption and is satisfied by, for example, mean-centered normal, $t$ and Gamma distributions.
Because $\theta_1$ and $\tilde{Y}_1$ are independent,  we have
\[
E( e^{ -\mathbf{i} k \theta_1} ) = E( e^{ - \mathbf{i} k  X_1}) [ E ( e^{ - \mathbf{i} k \tilde{Y}_1} ) ]^{-1}.
\]
Let $Y_1,\ldots, Y_m$ be as in (\ref{eq:1.8}). We observe that the $Y_l$'s are obtained from the voltage measurements at different time-points of the silent region of the spike train.
Then $Y_1, \ldots, Y_m$ are i.i.d.\ with
density $f_{0,\sigma^2}$ and are also independent of $X_1, \ldots, X_n$.
As $Y_1$ has the same distribution as $\tilde{Y}_1$, we have
\begin{equation}\label{eq:3.2}
E (e^{- \mathbf{i} k  Y_1}) \neq 0  \qquad  \forall k\in \{ 1, \ldots, p\}.
\end{equation}


Since $M_p = E[T_p( \theta)]$, we shall estimate $M_p$ using its sample analog, that is, the $(p+1)\times (p+1)$ matrix $\hat{M}_p$ whose $(j,k)$th element is given by
\begin{equation}\label{eq:3.1}
(\hat{M}_p )_{j,k} = \frac{ n^{-1} \sum_{i=1}^n e^{-\mathbf{i} (j-k) X_i} }{ m^{-1} \sum_{l =1}^m e^{-\mathbf{i} (j-k) Y_l}}.
\end{equation}
$\hat{M}_p$ is a Hermitian matrix and, hence, its eigenvalues are real numbers.
The method-of-moments estimator $\hat{\nu}$ for $\nu$ is as follows.
Let $\gamma$, $p_\gamma$ and $p\geq p_\gamma$ be as in Corollary \ref{cy:2.1}. Define
\begin{equation}\label{eq:3.3}
\hat{\nu} = \#\{ i\dvtx 1\leq i\leq p+1, \lambda_i (\hat{M}_p) > \gamma_{\mathrm{threshold}} \},
\end{equation}
where $\gamma_{\mathrm{threshold}} = \gamma \sqrt{p+1}$ and $\#\{\cdot\}$ denotes set cardinality.
We call $\gamma_{\mathrm{threshold}}$ the threshold parameter of $\hat{\nu}$.

\begin{tm} \label{tm:3.2}
Let $\hat{\nu}$ be as in (\ref{eq:3.3}) with $p\geq p_\gamma$. Suppose (\ref{eq:3.2}) and the conditions of Theorem \ref{tm:2.1} hold.
Then $\hat{\nu} \rightarrow \nu$ almost surely as $\min(m, n) \rightarrow \infty$.
\end{tm}

For $\hat{\nu}$ to perform well, it is necessary to obtain a good choice of the threshold parameter $\gamma_{\mathrm{threshold}}$.
The usual methods, such as cross-validation,  for determining the value of the threshold parameter do not seem to work here.
Instead we shall compute explicit error bounds for $\lambda_i (\hat{M}_p)$, $i=1,\ldots, p+1$, below.
These error bounds shall serve as guidelines for setting the value of $\gamma_{\mathrm{threshold}}$.

\begin{la} \label{la:3.1}
Let $\varepsilon>0$ be a constant,
\[
\Omega_{j,\varepsilon} = \biggl\{ \biggl| \frac{ m^{-1} \sum_{l =1}^m e^{-\mathbf{i} j Y_l} }{ E( e^{-\mathbf{i} j Y_1} )} -1\biggr| > \varepsilon \biggr\}\qquad   \forall j=1,\ldots, p,
\]
and $\Omega_\varepsilon = \bigcup_{j=1}^p \Omega_{j, \varepsilon}$.
Then
\begin{eqnarray*}
P( \Omega_\varepsilon ) &\leq& \sum_{j=1}^p \min  \biggl\{ \frac{6}{ m^2 [\varepsilon |E( e^{-\mathbf{i} j Y_1} )| ]^4} \biggl(1 +
O\biggl(\frac{1}{m}\biggr)\biggr),\\
&&
\phantom{\sum_{j=1}^p \min  \biggl\{}
\frac{ 42}{ m^3 [\varepsilon |E( e^{-\mathbf{i} j Y_1} )| ]^6}
\biggl(1 + O\biggl(\frac{1}{m}\biggr)\biggr)  \biggr\}.
\end{eqnarray*}
\end{la}

It is interesting to note that for the parameter values and sample sizes that we are concerned with in this article, the inequality of Lemma \ref{la:3.1} gives
a smaller upper bound than those obtained via the Hoeffding or Bernstein exponential-type inequalities.

\begin{tm} \label{tm:3.1}
Let $\Omega_\varepsilon$ be as in Lemma \ref{la:3.1}, $\Omega_\varepsilon^c$ be its complement and $E^{\Omega_\varepsilon^c}$
denote the conditional expectation given $\Omega_\varepsilon^c$. Then with the assumptions of Theorem \ref{tm:2.1},
\begin{eqnarray}\label{eq:3.4}
&&\sqrt{E^{\Omega_\varepsilon^c} \frac{\sum_{i=1}^{p+1} [ \lambda_i (\hat{M}_p ) - \lambda_i (M_p) ]^2 }{p+1}
}\nonumber\\[-8pt]\\[-8pt]
&&\qquad \leq
\sqrt{ \frac{2}{n (1 - \varepsilon)^2} \sum_{j=1}^p \frac{ p-j+1 }{ ( p+1) | \psi_Z(\sigma j)|^2 }  + \frac{ p\varepsilon^2 }{(1 - \varepsilon)^2} },
\nonumber
\end{eqnarray}
where $Z = Y_1/\sigma$, $\psi_Z( t) = E e^{-\mathbf{i} t Z}$ for all $t\in \mathbb R$ and, hence,
 $E e^{-\mathbf{i} j Y_1} = \psi_Z (\sigma j)$.
\end{tm}

We remark that the upper bounds of the inequalities of Lemma \ref{la:3.1} and Theorem~\ref{tm:3.1}, though relatively simple, are conservative in that the quantities on the left-hand side are substantially smaller than
those on the right-hand side.
Nonetheless, we shall end this section with an example which computes the upper bounds in Lemma~\ref{la:3.1} and Theorem \ref{tm:3.1} explicitly.
Suppose $\varepsilon = 0.05$, $p=20$, $\sigma = 0.1$, $Y_1 \sim N( 0, \sigma^2)$, $n=1000$ and $m= n^2$. Then
$P( \Omega_\varepsilon ) \leq 0.01$ and
\[
E^{\Omega_\varepsilon^c} \frac{\sum_{i=1}^{p+1} [ \lambda_i (\hat{M}_p ) - \lambda_i (M_p) ]^2 }{p+1}
\leq 0.12.
\]

\section{Simulated spike train study}\label{sec4}

In this section we shall study the finite sample performance of the method-of-moments estimator $\hat{\nu}$ given by (\ref{eq:3.3}) via simulated spike trains.

\begin{figure}[b]

\includegraphics{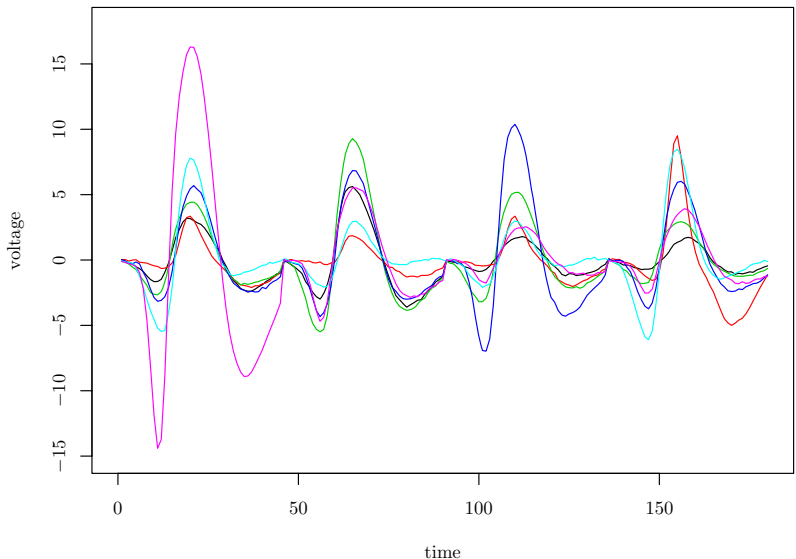}

  \caption{Six four-channel spike shapes from the locust data.}
  \label{fig:1}
\end{figure}

\subsection{Spike train generation}\label{sec4.1}

For realism, we use spike shapes estimated from real data to generate
the spike train. The spike shapes are obtained~by~applying Pouzat's
software SpikeOMatic which is available at
\href{http://www.biomedicale.univ-paris5.fr/SpikeOMatic.html}{www.biomedicale.}
\href{http://www.biomedicale.univ-paris5.fr/SpikeOMatic.html}{univ-paris5.fr/SpikeOMatic.html}
to a tetrode data set
recorded from the locust (\textit{Schistocerca americana}) antennal
lobe. This tetrode data is distributed with the SpikeOMatic software at
\href{http://www.biomedicale.univ-paris5.fr/SpikeOMatic/Data.html}{www.biomedicale.univ-paris5.fr/}
\href{http://www.biomedicale.univ-paris5.fr/SpikeOMatic/Data.html}{SpikeOMatic/Data.html}.
It is a 20-second
recording sampled at 15 kHz from a tetrode filtered between 300 Hz and
5 kHz. Pouzat, Mazor and Laurent (\citeyear{PML2002}) and Pouzat et al. (\citeyear{Petal2004}) are two
papers behind the software SpikeOMatic.

\begin{figure}

\includegraphics{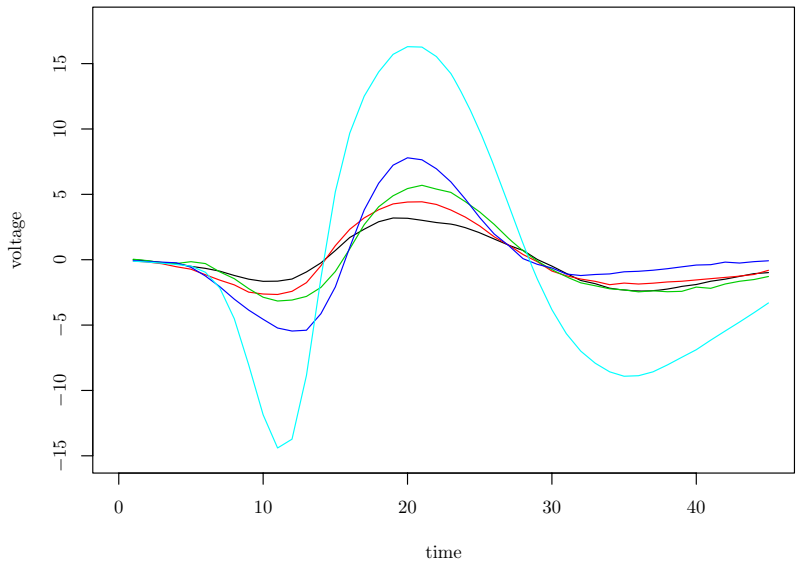}

  \caption{Five one-channel spike shapes selected with $\sigma=1.$}
  \label{fig:2}
\end{figure}

SpikeOMatic is applied to the locust data and we selected the 6 spike shapes with the largest numbers of spikes.
These (four-channel) spike shapes are shown in Figure \ref{fig:1}. As the two spike shapes with the smallest first channel positive peak heights have significant difference only in the fourth channel, we will delete one of them in order to generate single channel data in which distinct spike shapes are different enough to be distinguished. The resulting 5 spike shapes are shown in Figure \ref{fig:2}
with the noise standard deviation $\sigma = 1$.

Once the spike shapes are obtained, a Poisson process is simulated to select the time of spike events. Given
that a spike occurred, a random variable with categorical distribution is generated indicating which neuron has spiked.
The categorical distribution is assumed to have equal probabilities for each of $\nu$ possible spike shapes where $\nu$ is the number of neurons
(i.e.,\ $\pi_1=\cdots = \pi_\nu \leq 1/\nu$). Here $\nu$ ranges from 1 to 5.
The $\pi_i$'s can be set to different values but they should not be too close to zero.
This is necessary because if one of the neurons has infinitesimally small probability of spiking, then it would appear that no algorithm can estimate $\nu$ well with a finite sample.

As noted previously, Pouzat's locust data is a 20-second recording
sampled at 15~kHz. Thus, the recording has 300,000 time-points on a
regular grid. There are about 1000 spikes detected. Hence, the
collective firing rate is about $1000/300\mbox{,}000=1/300$. In our
simulations, we choose a similar collective firing rate of $1/400$, as
this gives about $10\%$ overlapping spikes.

SpikeOMatic rescales the data such that the noise standard deviation \mbox{$\sigma = 1$}.
In the generation of a spike train, independent standard Gaussian noise, or $t_5$ distributed noise (with degree of freedom 5) multiplied by $\sqrt{3/5}$,
is added to the ``signal.'' Multiplication by $\sqrt{3/5}$
 is needed since $t_5$ distribution does not have standard deviation 1.
Here ``signal'' refers to either the spike shape or 0 (in the case of pure noise).

\subsection{Spike detection}\label{sec4.2}

As mentioned in the Introduction, one of the tasks of spike sorting is to identify each spike in the spike train.
This is known as spike detection.
The function find.spikes.with.template provided by SpikeOMatic is used to detect spikes from spike train recordings.
In our study, the detection threshold is set to $2.25 \sigma$ (which is one of the recommended values of SpikeOMatic) where $\sigma$ is the noise standard deviation.
This is because the spike shape with the smallest positive peak height (in Figure \ref{fig:2}) is close to $3\sigma$ and
we do not want to consistently miss detecting spikes with this spike shape.
The function find.spikes.with.template uses template to detect spikes and, hence, there should not be too many false positives
with detection threshold $= 2.25 \sigma$ if the noise is  Gaussian. However, this threshold would present problems if the noise has a heavy-tailed
distribution such as the $t_5$ distribution (cf. Experiment \ref{exp6} of Section \ref{sec4.4}). For more details of spike detection via SpikeOMatic,
we refer the reader to the documentation for the software.

\subsection{Setting the tuning parameters of $\hat{\nu}$}\label{sec4.3}

We observe from the definition of $\hat{\nu}$ in (\ref{eq:3.3}) that $\hat{\nu}$ is not scale invariant and that
$\hat{\nu}$ has three tuning parameters, namely, $\sigma$, $p$ and $\gamma_{\mathrm{threshold}}$, to be determined in order to use $\hat{\nu}$.
First we observe that it is usually the case that the silent region forms a sizeable part of a spike train. Since data from the silent region can be used to estimate $\sigma$,
we shall, without loss of generality, assume in this subsection that $\sigma$ is known.

The function make.sweeps of SpikeOMatic extracts the spikes and the pure noise observations and automatically rescales them such that the
noise standard deviation is 1. After extracting the spikes, we shall further rescale them such that the noise standard deviation $\sigma=0.1$.
The rationale for this will be explained later.
Next we set $\gamma_{\mathrm{threshold}} = 1$ with the implicit assumption that $(p+1)\pi_\nu > \gamma_{\mathrm{threshold}}$. We observe from (\ref{eq:2.3}) that
the latter is a rough proxy for $\lambda_\nu (\hat{M}_p) > \gamma_{\mathrm{threshold}}$. This implicit assumption is indeed satisfied for all the 8 experiments.
(If this does not hold, we would have to set a smaller value for $\gamma_{\mathrm{threshold}}$ with a corresponding increase in false positives.)

Let $S_1,\ldots, S_n \in {\mathbb R}^d$ denote the extracted spikes.
The (normalized) first principal component $\alpha\in {\mathbb R}^d $
of $S_1,\ldots, S_n, 0, \ldots, 0$ is computed. Here $0\in {\mathbb
R}^d$ and the number of $0$'s used to compute $\alpha$ is taken to be
$0.01 n$ (to the nearest integer). The $0$'s are needed in the
computation of $\alpha$ because if not, $\alpha$ is not well defined in
the case of $\nu=1$ and i.i.d.\ Gaussian noise. Next define $X_i =
\alpha' S_i$ as in the \hyperref[sec1]{Introduction}. The rationale for projecting the
spikes onto the first principal component is the hope that this
direction will best separate the spike shapes from one another as well
as from pure noise (which is represented by $0$'s).


Now we argue that $\sigma$ should be neither too small nor too large after rescaling. If $\sigma$ is large, then $\psi_Z (\sigma j)$ will be small.
[For example,\ if the noise is Gaussian, we have $\psi_Z (\sigma j) =e^{-j^2\sigma^2/2}$.]
Thus, the bound as on the right-hand side of (\ref{eq:3.4}) is large.
This indicates that the estimation error of the eigenvalues will be large, resulting in poor performance of $\hat\nu$.

On the other hand, if $\sigma$ is scaled too small, the distance between some pair of $\mu_j$'s would likely be small,
that is,\ for some $j \neq k$, $|1-e^{\mathbf{i}(\mu_j-\mu_k)}|$ will be small. This implies that the lower bound in (\ref{eq:2.3})
can be less than the threshold parameter $\gamma_{\mathrm{threshold}} = 1$. This again results in poor performance of $\hat{\nu}$.
Consequently, we would like the  following condition to be satisfied:
\[
\mbox{\textsc{Condition (I)}:}\qquad    \sqrt{ \frac{2}{ 0.95^2 n} \sum_{j=1}^p \frac{ p-j+1 }{ ( p+1) | m^{-1}\sum_{l=1}^m e^{-\mathbf{i} p Y_l} |^2 }
+ \frac{0.05^2 p}{0.95^2}} \leq \frac{1}{3}.
\]


\noindent Since $m$ is large, we observe that $| E(e^{ -\mathbf{i} p Y_1})| \approx | m^{-1} \sum_{l=1}^m e^{ - \mathbf{i} p Y_l}|$.
Consequently, the left-hand side of the inequality in Condition (I) is an approximation for the right-hand side of (\ref{eq:3.4}) with $\varepsilon= 0.05$ and
can be used as an approximate upper bound for
the standard error of the eigenvalue estimates in (\ref{eq:3.4}). Thus, if Condition~(I) holds, $\gamma_{\mathrm{threshold}} = 1$ is very likely to  exceed
3 times this standard error and that,  for sufficiently  small $\pi_{\mathrm{cont}}$, $\lambda_{\nu+1} (\hat{M}_p)  \leq 1$.
%
Finally we choose $p = p_{\max}$ where $p_{\max}$ is the largest value of $p$ satisfying Condition~(I).
Numerical experiments on the accuracy of $\hat{\nu}$ with these values of $\sigma = 0.1$, $p = p_{\max}$, and $\gamma_{\mathrm{threshold}}=1$
will be reported in Section \ref{sec4.4}.

\subsection{Numerical experiments}\label{sec4.4}

In this subsection we shall study the performance of the estimator
$\hat{\nu}$ of Section \ref{sec4.3} via 8 simulated spike train
experiments. Each experiment is divided into 5 scenarios depending on
the number of neurons generating the spike train (i.e.,\ $\nu= 1 ,
\ldots, 5$). A total of at most 5 neurons are considered. The spike
shapes of these neurons are given in Figure \ref{fig:2}. As there are
many ways of selecting $\nu$ neurons from 5 neurons if $\nu<5$, we
shall choose the $\nu$ neurons in a ``least favorable'' manner for
estimating $\nu$: if $\nu =1$, we take the neuron with the smallest
peak height in Figure~\ref{fig:2} to be the one generating the spike
train; if $\nu=2$, then we take the 2 neurons with the two smallest
peak heights in Figure~\ref{fig:2} to be the  ones generating the spike
train; and so on until we reach $\nu=5$. In all the experiments the
estimator $\hat{\nu}$ is used after rescaling the data so that
$\sigma=0.1$ and setting $p= p_{\max}$ and
$\gamma_{\mathrm{threshold}}= 1$. The spike shape vectors each have
$d=45$ components. Table \ref{table:3.1.0} summarizes the 8 experiments
as a factorial design.

\begin{exper}\label{exp1}
In this experiment the noise is i.i.d.\ Gaussian with mean $0$ and variance $\sigma^2 = 1$.
We assume that the spikes are detected from the spike train without error (or, equivalently, there exists an oracle spike detector):
\begin{itemize}
\item The proportion of overlapping spikes $\pi_{\mathrm{cont}} \approx 0.1$ (or $10\%$).
\item $n=1000$ (or $500$), that is, there are $1000$ (or $500$) spikes detected from the spike train.

\item $m= 2 n$, where $m$ is the number of pure noise observations $Y_j$'s [as in (\ref{eq:1.8})] obtained from the silent region of the spike train.

\item The approximate $\mu_i$'s (i.e.,\ the projections of the spike
shape vector onto the first principal component $\alpha \in {\mathbb
R}^d$) of these neurons are presented in Table \ref{table:3.3.1}.
\end{itemize}
\end{exper}

\begin{table}
\caption{Approximate values of $\mu_1,\ldots,\mu_\nu$ for Experiments
\protect\ref{exp1}, \protect\ref{exp3}, \protect\ref{exp5} and \protect\ref{exp7}}\label{table:3.3.1}
\begin{tabular}{@{}lcd{2.1}d{2.1}ccc@{}}
\hline
 &   &  \multicolumn{1}{c}{$\bolds{\mu_1}$} & \multicolumn{1}{c}{$\bolds{\mu_2}$} & $\bolds{\mu_3}$ & $\bolds{\mu_4}$ & $\bolds{\mu_5}$\\
\hline
 Experiment \ref{exp1} & $\nu=1$ &  11.1  &   &   &  & \\
&  $\nu=2$ &  9.7  & 13.7  &   &  & \\
& $\nu=3$ &  8.4  &  11.3 & 15.8  &  & \\
&  $\nu=4$ &  5.7  & 9.5  & 12.3  & 20.4 & \\
&  $\nu=5$ &  11.4  & 14.3  &  16.9 & 19.5 & 58.9 \\[3pt]
 Experiment \ref{exp3}  & $\nu=1$ &  11.7  &   &   &  & \\
  & $\nu=2$ &  8.1  & 12.4  &   &  & \\
&  $\nu=3$ &  9.2  &  12.2 & 16.6  &  & \\
 & $\nu=4$ &  5.5  & 9.3  & 12.0  & 20.2 & \\
 & $\nu=5$ &  11.4  & 14.3  &  17.0 & 19.5 & 58.9 \\[3pt]
Experiment \ref{exp5}  & $\nu=1$ &  11.2  &   &   &  & \\
 & $\nu=2$ &  10.3  & 14.0  &   &  & \\
&   $\nu=3$ &  9.9  & 13.0 & 17.2  &  & \\
 & $\nu=4$ &  5.5  & 9.2  & 12.0  & 20.1 & \\
&  $\nu=5$ &  11.4  & 14.3  & 16.9 & 19.4 & 58.9 \\[3pt]
 Experiment \ref{exp7}  & $\nu=1$ &  11.6  &   &   &  & \\
 & $\nu=2$ &  8.5  & 12.7  &   &  & \\
&  $\nu=3$ &  9.1  & 12.2 & 16.5  &  & \\
 & $\nu=4$ &  5.5  & 9.2  & 12.0  & 20.2 & \\
 & $\nu=5$ &  11.4  & 14.3  & 17.0 & 19.6 & 58.9 \\
\hline
\end{tabular}
\end{table}


Table \ref{table:3.1.2} gives the percentage of the time (out of 100
repetitions) that $\hat{\nu} = \nu$, the true number of neurons
producing the spike train. In the case $n=1000$, $m=2000$,
$\hat{\nu}$ does well, for example, for $\nu=2$, the percentage of the
time $\hat{\nu}=\nu$ is $97\%$. In the case $n=500$ and $m=1000$,
$\hat{\nu}$ does well for the scenarios $1\leq \nu \leq 4$ but does
poorly for $\nu=5$.

\begin{remark*}
The Editor raised the following question: ``would detection improve if $m/n$ were larger than $2/1$?'' The nature in which the sample sizes $m,n$ affect
the accuracy of $\hat{\nu}$ is not a straightforward one. If $n$ is held fixed and $m$ is allowed to increase, then the accuracy of $\hat{\nu}$ should improve.
However, if $n$ is also allowed to increase, then $p_{\max}$ would likewise increase. This implies from (\ref{eq:3.4}) and Condition (I) that $\psi_Z (\sigma p_{\max})$
will be harder to estimate because the latter is further out in the tail of the characteristic function. Thus, in order to estimate $\psi_Z (\sigma p_{\max})$ accurately,
a much larger sample of $Y_1,\ldots, Y_m$ is needed.
Thus, doubling both sample sizes $m$ and $n$ may not necessarily increase the accuracy of $\hat{\nu}$. However, for sufficiently large $m$ (depending on the value of $n$),
the resulting $\hat{\nu}$ will improve in accuracy.
\end{remark*}

\begin{table}
\caption{Frequency (\%) of $\hat{\nu} =\nu$ with standard error in parentheses for Experiments
\protect\ref{exp1}--\protect\ref{exp8}}\label{table:3.1.2}
\begin{tabular*}{\textwidth}{@{\extracolsep{\fill}}lccccccc@{}}
\hline
  &   $\bolds{n}$ & $\bolds{m}$ & $\bolds{\nu=1}$ & $\bolds{\nu=2}$ & $\bolds{\nu=3}$ & $\bolds{\nu=4}$ & $\bolds{\nu=5}$\\
\hline
Experiment \ref{exp1} &  1000 & 2000 & 81 (3.9)  & \hphantom{0}97 (1.7) &  100 (0.0) & 100 (0.0) & 100 (0.0) \\
& \hphantom{1}500 &   1000 & 97 (1.7) & 100 (0.0) & 100 (0.0) & 100 (0.0) & \hphantom{0}10 (3.0) \\[3pt]
 Experiment \ref{exp2} & 1000 &  2000 & 51 (5.0)   &  \hphantom{0}79 (4.1) &   \hphantom{0}98 (1.4) &  100 (0.0) & 100 (0.0)\\
 & \hphantom{1}500 & 1000 & 65 (4.8) &  99 (1.0) & 100 (0.0) & 100 (0.0) & \hphantom{0}22 (4.1)\\[3pt]
Experiment \ref{exp3} & 1000 & 2000 & 89 (3.1)  & \hphantom{0}98 (1.4) &  100 (0.0) & 100 (0.0) & 100 (0.0)\\
 & \hphantom{1}500 & 1000 & 91 (2.9) & 100 (0.0) & 100 (0.0) & 100 (0.0) & \hphantom{00}5 (2.2) \\[3pt]
Experiment \ref{exp4} & 1000 & 2000 & 67 (4.7)  & \hphantom{0}98 (1.4) &  100 (0.0) & 100 (0.0) & 100 (0.0)\\
& \hphantom{1}500 & 1000 & 59 (4.9) & 100 (0.0) & 100 (0.0) & 100 (0.0) & \hphantom{0}21 (4.1) \\[3pt]
 Experiment \ref{exp5} & 1000 & 2000 & 82 (3.8)  & \hphantom{0}97 (1.7) &  100 (0.0) & 100 (0.0) & 100 (0.0)\\
& \hphantom{1}500 & $1000$  & 94 (2.4) & 99 (1.0) & 100 (0.0) & 100 (0.0) & \hphantom{00}6 (2.4) \\[3pt]
Experiment \ref{exp6} &  1000 & 2000 & \hphantom{0}0 (0.0)  & \hphantom{00}0 (0.0) &  \hphantom{0}99 (1.0) & 100 (0.0) & 100 (0.0)\\[3pt]
Experiment \ref{exp7} & 1000 & 2000 & 88 (3.2)  & 100 (0.0) &  100 (0.0) & 100 (0.0) & 100 (0.0)\\
& \hphantom{1}500 & $1000$  & 88 (3.2) & 100 (0.0) & 100 (0.0) & 100 (0.0) & \hphantom{00}8 (2.7) \\[3pt]
Experiment \ref{exp8} & 1000 & 2000 & \hphantom{0}0 (0.0)  & \hphantom{00}0 (0.0) &  100 (0.0) & 100 (0.0) & 100 (0.0)\\
\hline
\end{tabular*}
\end{table}

\begin{exper}\label{exp2}
This experiment is identical to Experiment \ref{exp1} except that, more
realistically, the oracle spike detector is not used. Instead, as
described in Section \ref{sec4.2}, we use the function
find.spikes.with.template (with spike detection threshold $2.25
\sigma$) provided by SpikeOMatic to detect the spikes of the spike
train. Table \ref{table:3.1.2} gives a summary of the percentage of the
time that $\hat{\nu} = \nu$, the true number of neurons producing the
spike train. We observe that for moderately large sample sizes
$n=1000$ and $m=2000$, the accuracy of $\hat{\nu}$ is still
reasonable though not as high as in Experiment \ref{exp1}. This is due
to the fact that spike detection is now not error free. We note that
the scenario $\nu=1$ with $\pi_{\mathrm{cont}} \approx 0.1$ is unlikely
to occur in practice due to the refractory period of a neuron which
prevents  the occurrence of overlapping spikes.


\begin{table}
\tablewidth=282pt
\caption{Frequency (\%) of $\hat{\nu}_1$ for Experiments \protect\ref{exp2} and
\protect\ref{exp3} with $n=1000$ and $m=2000$}\label{table:3.2.3}
\begin{tabular*}{282pt}{@{\extracolsep{\fill}}lccd{3.0}d{3.0}d{2.0}c@{}}
\hline
 & & $\bolds{\hat{\nu}_1 = 1}$ & \multicolumn{1}{c}{$\bolds{\hat{\nu}_1 = 2}$} & \multicolumn{1}{c}{$\bolds{\hat{\nu}_1 = 3}$} & \multicolumn{1}{c}{$\bolds{\hat{\nu}_1 = 4}$} & $\bolds{\hat{\nu}_1 = 5}$\\
\hline
Experiment \ref{exp2}  & $\nu = 1$    & 0 &  100 & 0  & 0  & 0 \\
& $\nu = 2$   & 0  & 100  & 0 & 0  & 0 \\
& $\nu = 3$  & 0 & 100 & 0 & 0 & 0 \\
& $\nu = 4$ & 0 & 20 & 80 & 0 & 0 \\
& $\nu = 5$  & 0 & 0 & 80 & 20 & 0 \\[6pt]
Experiment \ref{exp3}  & $\nu = 1$    & 0 &  100 & 0  & 0  & 0 \\
&  $\nu = 2$   & 0  & 100  & 0 & 0  & 0 \\
& $\nu = 3$  & 0 & 70 & 30 & 0 & 0 \\
& $\nu = 4$  & 0 & 0 & 100 & 0 & 0 \\
& $\nu = 5$  & 0 & 0 & 100 & 0 & 0 \\
\hline
\end{tabular*}
\end{table}

As a comparison, we shall now compute the SpikeOMatic estimate $\hat{\nu}_1$ of $\nu$. Briefly, the SpikeOMatic software uses an EM algorithm to compute $\hat{\nu}_1$
based on a penalized likelihood function. The likelihood is a finite mixture of multivariate normal distributions and the penalty is derived from the Bayesian information criterion (BIC).
This implies the assumption of Gaussian noise and no overlapping spikes. Another point of note is that the SpikeOMatic software assumes
at least 2 neurons generating the spike train and, hence, $\hat{\nu}_1$ is always  $\geq $2. There are 2 major tuning parameters for the SpikeOMatic software.
We take these to be nb.samples.per.site${}={}$3 and tolerance.factor${}={}$3.5. The other parameters are set as in the software tutorial 1 distributed with SpikeOMatic.
The performance of $\hat{\nu}_1$ appears to be rather robust to the choice of the 2 tuning parameters.
Table \ref{table:3.2.3} gives the frequency (\%) of $\hat{\nu}_1$ for 10 repetitions of the spike train.
In particular, when $\nu=2$, the frequency of $\hat{\nu}_1 = \nu$ is 100\% and when $\nu=3,4$ or $5$, the frequency of $\hat{\nu}_1 = \nu$ is 0\%.
\end{exper}

\begin{exper}\label{exp3}
In this experiment the noise is i.i.d.\ $N(0,1)$.
We assume that the spikes are detected from the spike train without error:
\begin{itemize}
\item The proportion of overlapping spikes $\pi_{\mathrm{cont}} = 0$ or, equivalently, that there are no overlapping spikes.

\item $n=1000$ (or $500$) and $m= 2 n$.

\item
The approximate $\mu_i$'s of these neurons are presented in Table \ref{table:3.3.1}.
\end{itemize}
\end{exper}

Table \ref{table:3.1.2} gives the percentage of the time (out of 100 repetitions) that $\hat{\nu} = \nu$.

As a comparison, we shall now compute the SpikeOMatic estimate $\hat{\nu}_1$ for $\nu$.
We take the major 2 tuning parameters to be nb.samples.per.site${}={}$3 and tolerance.factor${}={}$3.5. The other parameters are set as in the software tutorial 1 distributed with SpikeOMatic.
Table \ref{table:3.2.3} gives the frequency (\%) of $\hat{\nu}_1$ for 10 repetitions of the spike train.
In particular, when $\nu=2$, the frequency of $\hat{\nu}_1 = \nu$ is 100\%, when $\nu=3$, the frequency of $\hat{\nu}_1 = \nu$ is 30\% and when $\nu=4$ or $5$,
the frequency of $\hat{\nu}_1 = \nu$ is 0\%.

\begin{exper}\label{exp4}
This experiment is identical to Experiment \ref{exp3} except that
we use the function find.spikes.with.template (with spike detection threshold $2.25 \sigma$)
provided by SpikeOMatic to detect the spikes of the spike train. Table \ref{table:3.1.2}
gives the percentage of the time (out of 100 repetitions) that $\hat{\nu} = \nu$.
\end{exper}


\begin{exper}\label{exp5}
 In this experiment the noise has the $t_5$ distribution multiplied by $\sqrt{3/5}$
(so as to have mean $0$ and variance $1$).
We assume that the spikes are detected from the spike train without error:
\begin{itemize}
\item The proportion of overlapping spikes $\pi_{\mathrm{cont}} \approx 0.1$ (or $10\%$).
\item $n=1000$ (or $500$) and $m=2 n$.

\item
The approximate $\mu_i$'s (i.e.,\ the projections of the spike shape vector onto
the first principal component $\alpha \in {\mathbb R}^d$) of these neurons are presented in Table \ref{table:3.3.1}.
\end{itemize}
\end{exper}

Table \ref{table:3.1.2} gives the percentage of the time (out of 100 repetitions) that $\hat{\nu} = \nu$.

\begin{exper}\label{exp6}
This experiment is identical to Experiment \ref{exp5} except that
we use the function find.spikes.with.template (with spike detection threshold $2.25 \sigma$)
provided by SpikeOMatic to detect the spikes of the spike train. Table \ref{table:3.1.2}
gives the percentage of the time (out of 100 repetitions) that $\hat{\nu} = \nu$.
Here the accuracy of $\hat{\nu}$ is poor for $\nu=1$ or $2$. This is almost certainly due to the heavy-tailed $t_5$ distribution
presenting severe difficulties to the SpikeOMatic spike detection algorithm with threshold set to $2.25 \sigma$.
\end{exper}


\begin{exper}\label{exp7}
 In this experiment the noise has the $t_5$ distribution multiplied by $\sqrt{3/5}$
(so as to have mean $0$ and variance $1$).
We assume that the spikes are detected from the spike train without error:
\begin{itemize}
\item The proportion of overlapping spikes $\pi_{\mathrm{cont}} =0.0$, that is, there are no overlapping spikes.
\item $n=1000$ (or $500$) and $m=2 n$.

\item
The approximate $\mu_i$'s  of these neurons are presented in Table \ref{table:3.3.1}.
\end{itemize}
\end{exper}

Table \ref{table:3.1.2} gives the percentage of the time (out of 100 repetitions) that $\hat{\nu} = \nu$.

\begin{exper}\label{exp8}
This experiment is identical to Experiment \ref{exp7} except that
we use the function find.spikes.with.template (with spike detection threshold $2.25 \sigma$)
provided by SpikeOMatic to detect the spikes of the spike train. Table \ref{table:3.1.2}
gives the percentage of the time (out of 100 repetitions) that $\hat{\nu} = \nu$.
As in Experiment~\ref{exp6},  the accuracy of $\hat{\nu}$ here is poor for $\nu=1$ or $2$ and is due to the heavy-tailed $t_5$ distribution
presenting severe difficulties to the SpikeOMatic spike detection algorithm with threshold set to $2.25 \sigma$.
\end{exper}


\subsection{Study summary}\label{sec4.5}

The 8 experiments indicate that for moderately large sample sizes such
as $n=1000$ and $m=2000$, $\hat{\nu}$ is capable of very good
performance if the noise is i.i.d.\ Gaussian. If the noise has a
heavy-tailed distribution such as the $t_5$ distribution, $\hat{\nu}$
is still capable of very good performance if an oracle (error free)
spike detector is used. One reason for the good performance of
$\hat{\nu}$ for moderately large samples is that in all the experiments
the first principal component separates the spike shapes from one
another as well as from pure noise very well.

However, for smaller sample sizes such as $n = 500$ and $m = 1000$, $\hat{\nu}$ performs reasonably well for $1\leq \nu\leq 4$ if the noise is i.i.d.\ Gaussian but performs poorly for $\nu=5$.
This implies that the sample size $n=500$ is too small for $\hat{\nu}$ to handle the latter case well.

\section{A real data example}\label{sec5}

This section considers two spike train data sets. Both of these data sets are taken from Lewicki (\citeyear{L1994}).
One data set is an actual  40-second spike train recording. The other data set is a synthesized recording using 6 spike shapes estimated from the former data set.
For each of the 2 data sets, we shall compute and compare the estimates of the number of neurons generating the spike train
by applying Lewicki's software SUN Solaris OS version 1.1.8 [cf.\ Lewicki (\citeyear{L1994})], SpikeOMatic version 0.6-1 [cf.\ Pouzat, Mazor and Laurent (\citeyear{PML2002})]
and our proposed method-of-moments estimator $\hat{\nu}$.
For simplicity, let $\hat{\nu}_1$ denote the estimate of $\nu$ given by SpikeOMatic and $\hat{\nu}_2$ be corresponding to the estimate given by the Lewicki software.

\subsection{Synthesized data set}\label{sec5.1}

The spike shapes, the number of spikes from each neuron and the number of overlapping spikes for the synthesized recording can be found in Section 7 of
Lewicki (\citeyear{L1994}). The true number of neurons $\nu$ for the synthesized recording is 6. However, all three methods underestimate $\nu$.
 In particular, $\hat{\nu} = 4$ while $\hat{\nu}_1 = \hat{\nu}_2 =5$.
A likely reason is that the two smallest spike shapes are too close to each other for them to be identified as two distinct spike shapes (and not one) by the 3 estimators.
A graph of the 6 spike shapes is given in Figure 7 of Lewicki (\citeyear{L1994}).

With reference to $\hat{\nu}$, we have set the tuning parameters to $\gamma_{\mathrm{threshold}} = 1$, $\sigma= 0.1$ and $p= p_{\max}$ as in Section \ref{sec4}.
$p_{\max}$ turns out to be 18.
SpikeOMatic is used for spike detection and $n= 746$ spikes are detected.
Table \ref{table:2} lists the eigenvalues $\lambda_1 (\hat{M}_p) \geq \cdots \geq \lambda_{19} (\hat{M}_p)$ with
$\lambda_5(\hat{M}_p) = 0.96 < \gamma_{\mathrm{threshold}}=1$. Thus, as it stands, our estimate $\hat{\nu}$ misses returning the value 5 by a very narrow margin.
Also, we observe from Lewicki (\citeyear{L1994}) that $\pi_{\nu-1} + \pi_\nu \leq 32/895 \approx 0.036$, which is rather small. In fact, $(p_{\max} +1) 0.036 \approx 0.68$, which is way below
$\gamma_{\mathrm{threshold}}$. Thus, if we suspect that $\pi_\nu$ is this small,
we should lower the value of $\gamma_{\mathrm{threshold}}$ to below 1. For example, if $\gamma_{\mathrm{threshold}}$ is set to be 0.8
say,
 we shall obtain $\hat{\nu}= 5$ for the synthesized data set. However, reducing the value of $\gamma_{\mathrm{threshold}}$ would, of course, increase the chance of false positives too.

\subsection{Actual spike train data}

When the 3 methods are applied to the actual spike train data set, we obtain
$\hat{\nu} = 4$ while $\hat{\nu}_1 = 12$ and $\hat{\nu}_2 = 9$.
Lewicki [(\citeyear{L1994}), page 1020] inferred that the number of neurons $\nu$ generating  the actual spike train recording is still 6.
Thus, it would appear that $\hat{\nu}_1$ and $\hat{\nu}_2$ have overestimated $\nu$ and that they give rather different values of $\nu$ for the synthesized
and actual data sets.
On the other hand, $\hat{\nu}$, with the tuning parameters as in Section \ref{sec4},  gives $\hat{\nu}= 4$ as in the synthesized data set. This shows that $\hat{\nu}$
is rather stable and is probably less variable than $\hat{\nu}_1$ or $\hat{\nu}_2$. As in the discussion of Section \ref{sec5.1},
if we suspect that $\pi_\nu$ is very small such that $(p_{\max} +1) \pi_\nu < 1$, we should lower the value of $\gamma_{\mathrm{threshold}}$ in order to detect this spike shape.
If we reduce the value of $\gamma_{\mathrm{threshold}}$ to $0.8$, we obtain $\hat{\nu}= 5$ (from the eigenvalues of Table \ref{table:2}).

We conclude this section by  presenting the parameter settings for the
three methods. The parameters for Lewicki's software are set as his
default values for the analysis of the synthesized recording. Here a
spike is defined as the window of measurements that are within 1.0
millisecond prior to the occurrence of the peak and 4.0 milliseconds
after the occurrence of the peak. Since the sampling frequency is
20 kHz, there are 20 measurements before the peak and 80 measurements
after the peak.

The parameters for SpikeOMatic software are given in Table \ref{table:3}. For the detailed explanation of the meaning of the parameters, we refer the reader to the
SpikeOMatic manual which comes along with the software. We further note that the spike length is chosen to be $d=100$, the same as for Lewicki's software, and
$n=1447$ spikes are detected with $m=2000$.

With respect to $\hat{\nu}$, the spike detection procedure of
SpikeOMatic is used. This part of the parameter setting is the same as
in Table \ref{table:3}. Here $d=100$, spike detection threshold
$=3.00\sigma$ and the other tuning parameters for $\hat{\nu}$ are the
same as in Section \ref{sec4}.


\begin{table}
\tabcolsep=0pt
\caption{The eigenvalues in decreasing order computed from our algorithm for the two data sets}\label{table:2}
\begin{tabular*}{\textwidth}{@{\extracolsep{\fill}}lccccd{2.2}d{2.2}d{2.2}d{2.2}d{2.2}c@{}}
\hline
Synthesized recording
& 7.86 & 4.73 & 4.52 & 2.02 & 0.96 & 0.25 & 0.10 & 0.05 & 0.03 & 0.01\\
& 0.00 & 0.00 & 0.00 & 0.00 & 0.00 & 0.00 &-0.02 &-0.09 &-1.42 &\\[6pt]
Actual recording
& 6.64 & 3.83 & 1.59 & 1.17 & 0.86 & 0.67 & 0.48 & 0.33 & 0.20 & 0.14\\
& 0.11 & 0.07 & 0.05 & 0.01 &-0.04 &-0.11 & & & &\\
\hline
\end{tabular*}
\end{table}

\begin{table}
\tablewidth=130pt
\caption{Parameter settings for SpikeOMatic}\label{table:3}
\begin{tabular*}{130pt}{@{\extracolsep{\fill}}ld{4.1}@{}}
\hline
\textbf{Parameter} & \multicolumn{1}{c@{}}{\textbf{Value}}\\
\hline
template.length & 160\\
threshold & 3.0\\
sweep.length & 100\\
peak.location & 20\\
nb.noise.evt & 2000\\
nb.samples.per.site & 6 \\
tolerance.factor & 3.5 \\
nb.clusters.min & 2\\
nb.cluster.max & 12\\
nb.iterations & 25\\
nb.tries & 20\\
\hline
\end{tabular*}
\end{table}

\section{Concluding remarks}\label{sec6}

In conclusion, this article proposes a new estimator~$\hat{\nu}$ for estimating the number of neurons $\nu$ in a multi-neuronal spike train.
$\hat{\nu}$~has a number of advantages over alternative estimators for $\nu$ in the existing literature.
First, it is a method-of-moments estimator and uses trigonometric moment matrices in its construction
(unlike maximum likelihood estimators). As a result, the assumptions needed for $\hat{\nu}$ are minimal.  Indeed, the model (\ref{eq:1.2}) on which it is based
is a nonparametric mixture distribution. (\ref{eq:1.2}) takes explicitly into account the possibility of overlapping spikes while no parametric assumptions are
made on the  noise distribution or the overlapping spike distribution.

Second, we have managed to develop a rigorous nonasymptotic theory in support of $\hat{\nu}$. This theory is reasonably simple and transparent.
In particular, it shows that $\hat{\nu}$ is a strongly consistent
estimator of $\nu$ under mild conditions. Also, perhaps more importantly, the nonasymptotic error bounds of Theorems \ref{tm:2.1} and  \ref{tm:3.1}
provide us with a way of setting the tuning parameters $\gamma_{\mathrm{threshold}}$, $p$ and $\sigma$ so as to ensure that $\hat{\nu}$ performs well in practice.
The latter is further justified by applying $\hat{\nu}$ to a number of
spike train simulations in Section \ref{sec4} and to an actual spike train data set in Section \ref{sec5}.

Finally, we have assumed independent noise (i.e.,\ the $\eta_{i,k}$'s and $\eta_{l,j}^*$'s of Section~\ref{sec1}) throughout this article.
 If the noise is a  stationary and weakly dependent process, Theorem \ref{tm:3.2} still holds (i.e.,\ $\hat{\nu}$
is strongly consistent) as long as
\begin{eqnarray}\label{eq:6.1}
\frac{1}{n} \sum_{i=1}^n e^{-\mathbf{i} (j-k) X_i} &\rightarrow & E e^{-\mathbf{i} (j-k) X_1},
\nonumber\\[-8pt]\\[-8pt]
\frac{1}{m} \sum_{l=1}^m e^{-\mathbf{i} (j-k) Y_l} &\rightarrow & E e^{-\mathbf{i} (j-k) Y_1}\qquad  \forall 1\leq j,k\leq p+1,
\nonumber
\end{eqnarray}
almost surely as $\min(m, n)\rightarrow \infty$. We observe that (\ref{eq:6.1}) is a rather mild condition and is satisfied by many weakly
dependent processes.

\begin{appendix}
\section*{Appendix}\label{app}
\begin{pf*}{Proof of Theorem \ref{tm:2.1}}
First suppose that $i \in \{1,\ldots, \nu \}$.
We observe from Lemma \ref{la:a.2} below that
$\lambda_i (M_{p, \mathrm{disc}} ) = \lambda_i (B)$. Let $B^\dag = \operatorname{diag}( (p+1)\pi_1,\ldots, (p+1)\pi_\nu)$.
Using Theorem A.37 of Bai and Silverstein (\citeyear{BS2009}), we have
\begin{eqnarray*}\label{eq:asymerr}
\sum_{i=1}^{\nu} [ \lambda_i( B)-\lambda_i( B^\dag) ]^2
&\le& \sum_{j=1}^{\nu}\sum_{k=1}^{\nu} | B_{jk}- B^\dag_{jk} |^2 \\
&=& 2\sum_{1\le j<k\le\nu}\pi_j\pi_k  \biggl|\frac{1-e^{\mathbf{i}(p+1)(\mu_j-\mu_k)}}{1-e^{\mathbf{i}(\mu_j-\mu_k)}} \biggr|^2.
\end{eqnarray*}
Thus,
\[
|\lambda_i( M_{p, \mathrm{disc}})- (p+1) \pi_i| \leq
\sqrt{2\sum_{1\le j<k\le\nu}\pi_j\pi_k  \biggl|\frac{1-e^{\mathbf{i}(p+1)(\mu_j-\mu_k)}}{1-e^{\mathbf{i}(\mu_j-\mu_k)}} \biggr|^2}.
\]
Since $M_p= M_{p, \mathrm{disc}} + \pi_{\mathrm{cont}} M_{p, \mathrm{cont}}$, we observe from Corollary 4.9 of Stewart and Sun (\citeyear{SS1990}) and Lemma \ref{la:a.3} that
\begin{eqnarray*}
\lambda_i( M_p)
&\geq & \lambda_i( M_{p, \mathrm{disc}})+\pi_{\mathrm{cont}} \lambda_{p+1}( M_{p, \mathrm{cont}})\\
&\geq & (p+1) \pi_i +2\pi\pi_{\mathrm{cont}} \Biggl\{ \min_{0\leq \mu < 2\pi} \sum_{j=-\infty}^\infty f_{\mathrm{cont}} (\mu+2\pi j) \Biggr\}
\\&&{}
 -\sqrt{2\sum_{1\le j<k\le\nu}\pi_j\pi_k  \biggl|\frac{1-e^{\mathbf{i}(p+1)(\mu_j-\mu_k)}}{1-e^{\mathbf{i}(\mu_j-\mu_k)}} \biggr|^2},
\\
\lambda_i( M_p )
&\leq & \lambda_i( M_{p, \mathrm{disc}})+\pi_{\mathrm{cont}} \lambda_1( M_{p,\mathrm{cont}})\\
&\le& (p+1) \pi_i +2\pi\pi_{\mathrm{cont}} \Biggl\{ \max_{0\leq \mu < 2\pi} \sum_{j=-\infty}^\infty f_{\mathrm{cont}} (\mu+2\pi j) \Biggr\}
\\&&{}
+\sqrt{2\sum_{1\le j<k\le\nu}\pi_j\pi_k  \biggl|\frac{1-e^{\mathbf{i}(p+1)(\mu_j-\mu_k)}}{1-e^{\mathbf{i}(\mu_j-\mu_k)}} \biggr|^2}.
\end{eqnarray*}
This proves the first statement of Theorem \ref{tm:2.1}.
Next, suppose that $i \in \{ \nu+1,\ldots, p+1\}$. Then $\lambda_i (M_{p,\mathrm{disc}}) =0$. Using Corollary 4.9 of Stewart and Sun (\citeyear{SS1990}) and Lemma \ref{la:a.3} again, we obtain
the second statement of Theorem \ref{tm:2.1}.
\end{pf*}

\begin{la} \label{la:a.1}
Let $m>n$ be positive integers and $A$ be a $m\times n$ matrix with complex-valued entries. Then the eigenvalues of $A A^*$ are the eigenvalues of $A^* A$ and $(m-n)$ zeros
where $A^*$ denotes the conjugate transpose of $A$.
\end{la}

\begin{pf}
We observe from the singular value decomposition of $A$ that
$A = U D V^*$ where $U$ is a $m\times m$ unitary matrix, $D$ a $m\times n$ diagonal matrix with nonnegative real numbers on the diagonal and $V$ a $n\times n$ unitary matrix.
Then
$A^* A = V D^* D V^*,$
and
$A A^* = U D D^* U^*$. Lemma \ref{la:a.1} follows since $D^* D$ and $D D^*$ are both diagonal matrices.
\end{pf}

\begin{la} \label{la:a.2}
Let $M_{p,\mathrm{disc}}$ be as in (\ref{eq:2.1}).
With the notation and assumptions of Theorem \ref{tm:2.1}, we have
\begin{eqnarray*}
\lambda_i (M_{p,\mathrm{disc}}) &=& \lambda_i (B)\qquad  \forall i=1,\ldots, \nu,
\nonumber \\
\lambda_i (M_{p,\mathrm{disc}}) &=& 0\qquad \hspace*{20.5pt}  \forall i= \nu+1,\ldots, p+1,
\end{eqnarray*}
where $B$ is a $\nu \times \nu$ Hermitian matrix defined by
\begin{eqnarray*}
B &=&  \left(\matrix{
(p+1) \pi_1 & \displaystyle\sqrt{\pi_1\pi_2}\frac{1-e^{\mathbf{i}(p+1)(\mu_1-\mu_2)}}{1-e^{\mathbf{i}(\mu_1-\mu_2)}} \cr
\displaystyle\sqrt{\pi_1\pi_2}\frac{1-e^{\mathbf{i}(p+1)(\mu_2-\mu_1)}}{1-e^{\mathbf{i}(\mu_2-\mu_1)}} & (p+1) \pi_2
\cr
\vdots & \vdots \vspace*{7pt}\cr
\displaystyle\sqrt{\pi_1\pi_\nu}\frac{1-e^{\mathbf{i}(p+1)(\mu_\nu-\mu_1)}}{1-e^{\mathbf{i}(\mu_\nu-\mu_1)}} &
\displaystyle\sqrt{\pi_2\pi_\nu}\frac{1-e^{\mathbf{i}(p+1)(\mu_\nu-\mu_2)}}{1-e^{\mathbf{i}(\mu_\nu-\mu_2)}}}\right.\\
&&\hspace*{110pt}
\left.\matrix{\ldots & \displaystyle\sqrt{\pi_1\pi_\nu}\frac{1-e^{\mathbf{i}(p+1)(\mu_1-\mu_\nu)}}{1-e^{\mathbf{i}(\mu_1-\mu_\nu)}}\cr
\ldots &
\displaystyle\sqrt{\pi_2\pi_\nu}\frac{1-e^{\mathbf{i}(p+1)(\mu_2-\mu_\nu)}}{1-e^{\mathbf{i}(\mu_2-\mu_\nu)}}\cr
 \ddots & \vdots\cr
 \ldots & (p+1) \pi_\nu}\right).
\end{eqnarray*}
\end{la}

\begin{pf}
Let
\[
\Xi =  \pmatrix{
1 & \ldots & 1\cr
e^{-\mathbf{i}\mu_1} & \ldots & e^{-\mathbf{i}\mu_\nu}\cr
e^{-\mathbf{i}2\mu_1}& \ldots & e^{-\mathbf{i}2\mu_\nu}\cr
\vdots & \ddots & \vdots \cr
e^{-\mathbf{i}p\mu_1} & \ldots & e^{-\mathbf{i}p\mu_\nu}
}
\pmatrix{
\sqrt{\pi_1} & 0 & \ldots & 0 \cr
0 & \sqrt{\pi_2} & \ldots & 0 \cr
\vdots & \vdots & \ddots & \vdots \cr
0 & 0 & \ldots & \sqrt{\pi_\nu}
}.
\]
Then $M_{p, \mathrm{disc}} = \Xi \Xi^*$ and $B= \Xi^* \Xi$. Lemma \ref{la:a.2} follows from Lemma \ref{la:a.1}.
\end{pf}

\begin{la} \label{la:a.3}
Let $M_{p, \mathrm{cont}}$ be as in (\ref{eq:2.1}).
With the notation and assumptions of Theorem \ref{tm:2.1}, we have
\begin{eqnarray*}
&&2\pi \min_{0\leq \mu < 2\pi} \sum_{j=-\infty}^\infty f_{\mathrm{cont}} (\mu+2\pi j)\\
&&\qquad \leq
 \lambda_{p+1}( M_{p, \mathrm{cont}} )\leq \lambda_1( M_{p, \mathrm{cont}})\\
 &&\qquad  \leq 2\pi \max_{0\leq \mu < 2\pi} \sum_{j=-\infty}^\infty f_{\mathrm{cont}} (\mu+2\pi j).
\end{eqnarray*}
\end{la}

\begin{pf}
Let $a=(a_1,\ldots,a_{p+1})' \in {\mathbb C}^{p+1}$. Then
\begin{eqnarray*}
\frac{a^* M_{p, \mathrm{cont}} a}{a^* a}
&=& \frac{\sum_{k=1}^{p+1} \sum_{j=1}^{p+1} [ \int_0^{2\pi} \sum_{l=-\infty}^\infty f_{\mathrm{cont}} (\mu+2\pi l) e^{\mathbf{i}(k-j)\mu} \,d\mu  ] a_k a_j^*}{a^* a}\\
&=& \frac{\int_0^{2\pi} |\sum_{k=1}^{p+1} a_k e^{ \mathbf{i} k\mu}|^2 \sum_{l=-\infty}^\infty f_{\mathrm{cont}} (\mu+2\pi l)\, d \mu }{\sum_k |a_k|^2}\\
&=& \frac{ 2 \pi \int_0^{2\pi} |\sum_{k=1}^{p+1} a_k e^{ \mathbf{i} k\mu}|^2 \sum_{l=-\infty}^\infty f_{\mathrm{cont}} (\mu+2\pi l) \,d \mu }{
\int_0^{2\pi} |\sum_k a_k e^{ \mathbf{i} k\mu}|^2 \,d\mu}.
\end{eqnarray*}
Thus, for an arbitrary $a \in {\mathbb C}^{p+1}$ such that $a^* a = 1$,
\begin{eqnarray*}
&&2\pi \min_{0\leq \mu < 2\pi} \sum_{j=-\infty}^\infty f_{\mathrm{cont}} (\mu+2\pi j) \\
&&\qquad \leq a^* M_{p, \mathrm{cont}} a\\
&&\qquad \leq 2\pi \max_{0\leq \mu < 2\pi} \sum_{j=-\infty}^\infty f_{\mathrm{cont}} (\mu+2\pi j).
\end{eqnarray*}
Since $\lambda_1 (M_{p, \mathrm{cont}}) = \sup_{a\in {\mathbb C}^{p+1}, a^* a =1} a^* M_{p, \mathrm{cont}} a$ and
$\lambda_{p+1} (M_{p, \mathrm{cont}}) = \break\inf_{a\in {\mathbb C}^{p+1}, a^* a =1} a^* M_{p, \mathrm{cont}} a$,
Lemma \ref{la:a.3} is proved.
\end{pf}

\begin{pf*}{Proof of Theorem \ref{tm:3.2}}
We observe from the strong law of large numbers that
$\hat{M}_p \rightarrow M_p$ almost surely as $\min(m, n)\rightarrow \infty$. This implies that
$\lambda_{\nu+1} (\hat{M}_p) < \gamma \sqrt{p+1} < \lambda_\nu (\hat{M}_p)$  almost surely as $\min(m,n)\rightarrow \infty$.
\end{pf*}

\begin{pf*}{Proof of Lemma \ref{la:3.1}}
Using Markov's inequality, we observe for $\ell = 1,\ldots, p$ that
\begin{eqnarray*}
P( \Omega_{\ell,\varepsilon} ) &=&
P \Biggl( \Biggl| \frac{1}{m} \sum_{j=1}^m ( e^{-\mathbf{i} \ell Y_j}  - E e^{-\mathbf{i} \ell Y_1} ) \Biggr| \geq \varepsilon |E( e^{-\mathbf{i} \ell Y_1} )| \Biggr)
\nonumber \\
&\leq &  \frac{ E[ | m^{-1} \sum_{j=1}^m ( e^{-\mathbf{i} \ell Y_j}  - E e^{-\mathbf{i} \ell Y_1} ) |^k ]}{ [\varepsilon |E( e^{-\mathbf{i} \ell Y_1} )| ]^k }\qquad
 \forall k\in {\mathbb Z}^+.
\end{eqnarray*}
Taking $k=3, 4$, we obtain
\begin{eqnarray*}
P( \Omega_{\ell,\varepsilon} ) &\leq& \min\biggl\{ \frac{6}{ m^2 [\varepsilon |E( e^{-\mathbf{i} \ell Y_1} )| ]^4} \biggl(1 +
O\biggl(\frac{1}{m}\biggr)\biggr),\\
&&\hphantom{\min\biggl\{}
\frac{ 42}{ m^3 [\varepsilon |E( e^{-\mathbf{i} \ell Y_1} )| ]^6}
\biggl(1 + O\biggl(\frac{1}{m}\biggr)\biggr) \biggr\}.
\end{eqnarray*}
\upqed
\end{pf*}

\begin{pf*}{Proof of Theorem \ref{tm:3.1}}
Using Theorem A.37 of Bai and Silverstein (\citeyear{BS2009}), we have
\begin{eqnarray}\label{eq:a.10}
&& E^{\Omega^c} \sum_{i=1}^{p+1} [ \lambda_i( \hat{M}_p)-\lambda_i( M_p) ]^2
\nonumber \\
&&\qquad \leq E^{\Omega^c} \sum_{j=1}^{p+1}\sum_{k=1}^{p+1} | (\hat{M}_p)_{j,k}- (M_p)_{j,k} |^2
\nonumber \\
&&\qquad = 2 \sum_{j=1}^p (p-j+1) E^{\Omega^c} \biggl| \frac{n^{-1} \sum_{i=1}^n e^{-\mathbf{i} j X_i} }{  m^{-1} \sum_{l =1}^m e^{-\mathbf{i} j Y_l}}
-  \frac{ E (e^{-\mathbf{i} j X_1} )}{ E( e^{-\mathbf{i} j Y_1}) } \biggr|^2
\nonumber \\
&&\qquad = 2 \sum_{j=1}^p (p-j+1)   \biggl[ E^{\Omega^c}  \biggl| \frac{n^{-1} \sum_{i=1}^n e^{-\mathbf{i} j X_i} }{  m^{-1} \sum_{l =1}^m e^{-\mathbf{i} j Y_l}}
-  \frac{ E (e^{-\mathbf{i} j X_1} )}{  m^{-1} \sum_{l =1}^m e^{-\mathbf{i} j Y_l} }  \biggr|^2
\nonumber \\
&&\qquad \quad\hspace*{96pt} {}  + E^{\Omega^c}  \biggl| \frac{E( e^{-\mathbf{i} j X_1}) }{  m^{-1} \sum_{l =1}^m e^{-\mathbf{i} j Y_l}}
-  \frac{ E (e^{-\mathbf{i} j X_1} )}{ E( e^{-\mathbf{i} j Y_1}) }  \biggr|^2  \biggr]
\nonumber\\[-8pt]\\[-8pt]
&&\qquad = 2 \sum_{j=1}^p (p-j+1)   \biggl[ E^{\Omega^c}  \biggl| \frac{n^{-1} \sum_{i=1}^n e^{-\mathbf{i} j X_i} }{  m^{-1} \sum_{l =1}^m e^{-\mathbf{i} j Y_l}}
-  \frac{ E (e^{-\mathbf{i} j X_1} )}{  m^{-1} \sum_{l =1}^m e^{-\mathbf{i} j Y_l} }  \biggr|^2
\nonumber \\
&&\qquad \hspace*{40pt}\quad {}  + |E (e^{-\mathbf{i} j X_1} )|^2
E^{\Omega^c}  \biggl| \frac{1 }{  m^{-1} \sum_{l =1}^m e^{-\mathbf{i} j Y_l}}  -  \frac{ 1}{ E( e^{-\mathbf{i} j Y_1}) }  \biggr|^2  \biggr]
\nonumber \\
&&\qquad \leq 2 \sum_{j=1}^p (p-j+1)   \biggl[ \frac{ E |n^{-1} \sum_{i=1}^n e^{-\mathbf{i} j X_i} - E (e^{-\mathbf{i} j X_1}) |^2 }{ (1 - \varepsilon)^2 |E (e^{-\mathbf{i} j
Y_1})|^2}\nonumber\\
&&\qquad \quad \hspace*{126pt}
{}+ \frac{ \varepsilon^2 |E (e^{-\mathbf{i} j X_1} )|^2}{ (1 - \varepsilon)^2 | E( e^{-\mathbf{i} j Y_1}) |^2 }  \biggr]
\nonumber \\
&&\qquad \leq \frac{2}{(1 - \varepsilon)^2} \sum_{j=1}^p (p-j+1) \biggl[ \frac{1}{n |E(e^{-\mathbf{i} j Y_1})|^2} + \varepsilon^2\biggr].\nonumber
\end{eqnarray}\upqed
\end{pf*}
\end{appendix}

\section*{Acknowledgments}

Wei-Liem Loh would like to thank Professors Rob Kass, Shinsuke Koyama
and Matt Harrison for the many discussions on neural spikes trains when
he visited the Department of Statistics, Carnegie Mellon University in
Fall 2006 and Summer 2007. We would also like to thank the Editor,
Professor Karen Kafadar and an Associate Editor for their suggestions
and comments that motivated Sections \ref{sec5} and \ref{sec4}, respectively.


\printaddresses

\end{document}